# A Kantian Solution for the Freedom of Choice Loophole in Bell Experiments

Romeu Rossi Júnior
Patrícia Kauark-Leite[1]

**Abstract**

Bell's theorem is based on three assumptions: realism, locality, and measurement independence. The third assumption is identified by Bell as linked to the freedom of choice hypothesis. He holds that ultimately the human free will can ensure the measurement independence assumption. The incomplete experimental conditions for supporting this third assumption are known in the literature as "freedom-of-choice loophole" (FOCL). In a recent publication, Abellán et al [2018] address this problem and follow this same strategy embraced by Bell [2004]. Nevertheless, the possibility of human freedom of choice has been a matter of philosophical debate for more than 2000 years, and there is no consensus among philosophers on this topic. If human choice is not free, Bell's solution would not be sufficient to close FOCL. Therefore, in order to support the basic assumption of this experiment, it is necessary to argue that human choice is indeed free. In this paper, we present a Kantian position on this topic and defend the view that this philosophical position is the best way to ensure that BigBell Test (Abellán et al. [2018]) can in fact close the loophole.

**Keywords:** Bell's inequality, free will, Kant, agency

## 1. Introduction

Performing a Bell test is not a simple task. Bell's theorem (Bell [1964]) is intended to show that local realist theories are not adequate for describing quantum phenomena. To reach an experimental result that allows such broad conclusions, it is necessary that the experiments rigorously fulfill two technical conditions: (i) speed and efficiency in the detection of measures and (ii) the permanence of entangled states that maintain their coherence even when separated at great distances. In addition, it is also necessary to presuppose a third condition, which is the unpredictable measurement settings. This third condition requires that the measurements settings imply random choices that should be statistically independent of any influence of hidden variables (Abellán et al [2018]). The first experimental implementations (Aspect, Grangier and Roger [1982]; Aspect,

[1] This work has received funding from the *Brazilian* Federal Agency for Support and Evaluation of Graduate Education (CAPES) and is part of the activities of *Kant in South America* (KANTINSA) Project, founding by Minas Gerais *Research Foundation* (FAPEMIG) *and the* European Union's Horizon 2020 research and innovation programme under the Marie Skłodowska-Curie grant agreement No 777786.

Dalibard, and Roger [1982]; Weihs et al [1998]; Ou and Mandel [1988]; Shih and Alley [1988]; Tapster et al [1994]) show agreement with quantum theory, but were not able to ensure the necessary preconditions for the tests. In these cases, the Bell test goal has not been not reached, since the same experimental results can be explained by classical theories using local hidden variables. This circumstances of incomplete experimental conditions are referred in the literature as "loopholes." There are different kinds of loopholes, most of them are related to technical issues. The detection efficiency loophole and the locality loophole were recently overcome in experimental settings (Hensen et al [2015]; Giustina et al [2015]; Shalm et al [2015]).

The freedom-of-choice loophole (FOCL) is related to the assumption that the measurement setting variables (**x** and **y**) are statistically independent of the hidden variables. Usually, in Bell tests, the measurement settings are determined by a device, a physical randomizer (Hensen et al [2015]; Rosenfeld et al [2017]; Abellán et al [2015]; Fürst et al [2010]; Scheidla et al [2010]) that would be responsible for the required statistical independence. However, in classical terms, we can say that the randomizers have a causal past because they are physical devices, and consequently the setting variables **x** and **y** have a common causal past. In other words, we can say that the backward light cone of the setting variables **x** and **y** overlaps. The fact that the measurement setting variables **x** and **y** have a causal past in common, entitles one to assume that a hidden variable can be a common cause for them. To ignore this possibility would be to assume that "Bell's theorem applies only to a hybrid universe in which hidden variables determine only part of the outcomes of experiments." (Brans, C. [1988]). This means that hidden variables would determine the measurable outcomes (**A** and **B**), but not the setting variables **x** and **y**.

Efforts to close FOCL have been made by different experimental groups. Scheidla et al. [2010] and Shalm et al, [2015] claim that they have fulfilled this task by ensuring that the measurement setting choices are separated in a spacelike way from the event of the creation of the particles. They made the strong assumption that the hidden variable ($\lambda$) is created together with the entangled pair of particles. Therefore, the spacelike separation between the particle pair creation event and the measurement setting choices of the variables **x** and **y** would ensure the statistical independence between **x** and **y**, on one side, and $\lambda$, on the other.

There are also attempts to tighten FOCL by constraining the space-time volume. In these attempts the possible causal relationship between the measurement setting variables (**x** and **y**) and the hidden variable ($\lambda$) could have occurred (Rauch et al [2018]; Handsteiner et al [2017]) only in a very remote past. This is done through experimental setups that allow the measurement setting choices to be made through degrees of freedom of photons emitted by distant stars. In this way, the experiment "pushes the origin of the measurement settings considerably deeper into cosmic history" (Rauch et al [2018]). According to Rauch et al [2018], only **4%** of the space-time volume is left available for establishing the causal relationship between the settings variables **x** and **y,** and λ.

However, Abellán et al [2018] hold that even the strong restrictions established by Dominik Rauch et al. [2018] would not be sufficient to close FOCL, since in all these attempts, the measurement settings continue to be defined by physical objects, as in the case of photons emitted by distant stars. In their words, "while still requiring a physical assumption, and thus not closing the FOCL, this strategy tightens the loophole in various ways" (Abellán et al [2018]). The assumption taken by Abellán et al [2018] is that the measurement settings can be considered legitimately independent of λ only if they are the result of human choices. For only in this situation is the link between the measurable variables and λ excluded. This idea goes back to the famous remark made by Bell in the last section of his article "The theory of local beables" (Bell [2004]):

> It has been assumed (in deriving Bell's theorem) that the settings of instruments are in some sense free variables - say at the whim of experimenters - or in any case not determined in the overlap of the backward light cones.

Bell and the authors of Big Bell Test (Abellán et al [2018]) support the idea that human choices are free, therefore, suitable to close the FOCL. Therefore, in the Big Bell Test, since the measurement setting variables in the tests were defined by human choices, the authors claim to have closed the FOCL.

By accepting the free will choice as the only way to close the FOCL, the authors of Big Bell test implicitly introduce into the process a different kind of causality, aka causality through freedom, that cannot be assimilated to the natural causal chain (natural causality). In philosophical terms, they tacitly admit that physical events can be caused either by natural causes or by human freedom. In the first case, the effect has a cause that was caused by another cause. In the second case, the effect has a cause that was not caused by any empirical cause and, in this sense, one can say that this cause is free. Thus, according to this view, the variable λ could be considered as a possible cause of any physical event, e.g., processes in random number generators, emission of photons in stars, etc., but not a cause of human choices. This philosophical thesis was admitted by Bell and by the authors of Big Bell test to be sufficient close the FOCL, and so to ensure the conditions of validity of Bell's theorem. At this point, it becomes clear that the initial expectation of transforming the philosophical debate between Einstein and Bohr on the incompleteness of quantum theory into a purely experimental investigation was not accomplished by Bell.

The philosophical debate on free will can be traced back to Augustine [2011], for whom the possibility of human freedom in presence of the foreknowledge of God was a central issue. In the XVIII century, with the advent of Newtonian physics, this debate acquires new contours. The free will thesis seemed to be opposed to the determinism of the laws of physics. In this sense, human freedom would not be possible if every event is determined by the previous state of things according to strict laws of nature. The philosophers who hold that universal natural determinism and freedom are mutually metaphysically inconsistent are classified as incompatibilists, whereas those who deny this possibility and assert the consistency of universal natural determinism and freedom are known as compatibilists.

However, the simple negation of determinism is not enough to hold a coherent position in favor of the free will assumption. If every event were to result from a random process, then human choices would have to result from a random process too, and thus they would have no agential causal source. If human choices have no agential causal source, they cannot be effect of human free will. Thus, in a universally indeterministic world, decisions and choices would not be caused by human free will because they would be random events and not have an agential source. Therefore, indeterminism is also a threat to human free will. Nevertheless, the case of "superindeterminism," i.e., universal natural indeterminism, which deny both the determinism and the causality, is excluded in the context of Bell experiments, which try to show precisely the inadequacy of causal theories applied to quantum phenomena, whether they are deterministic or probabilistic.

Conversely, if human choices were to be completely determined by natural causality, then free will would be an illusion. However, it is not necessary to assume that this natural causality is deterministic. We can postulate a kind of "compatibilistic incompatibilism" between natural probabilistic causality (not deterministic) and free will. The possibility of human choice being determined by a natural causality (not necessarily deterministic) is often identified as a "superdeterministic" position (Abellán et al [2018]). Nevertheless, Bell and the authors of Big Bell test hold that any kind of superdeterministic position cannot be experimentally tested. In their words: "the theory that the entire experiment, including choices and outcomes, is pre-determined by initial conditions is known as superdeterminism [and] superdeterminism cannot be tested" (Abellán et al [2018]). Thus, they assume in the experiment, although not explicitly, a philosophical position according to which freedom and nature belongs to two different domains. The first one is essentially human and agential, and the second one essentially physical and non-agential. It is only in the domain of essentially physical events that the concept of natural causality can be applied. In the domain of human action, the choices are free.

In this paper, we will show that the tacit philosophical commitment of Big Bell experiment is better understood in Kantian terms. We will therefore make explicit the Kantian philosophical basis of the central hypothesis of the experiment carried out by Abellán et al [2018], according to which human choices are free. For that, we will take into account a certain interpretation of Kant's theory of human agency defended among others by Henry Allison [1990] and Maria Borges [2019]. It is not the purpose of this work to go into detail concerning the discussions between Kantian scholars on the problem of the free will. Our aim is just to show that Big Bell experiment's account is consistent to the Kantian interpretation that attributes to the agent the power to start a causal series spontaneously, free from the influences of any previous events. In Kant's philosophy, natural causality allows us to understand the nature as subject to laws which hold universally in nature and give it a predictable character. In this sense, every event is preceded by a cause which generates a causal chain in accordance with a rule. However, that is not the only kind of causality that is operating in the empirical world. We need to presuppose a causality through freedom according to which the power of choice of the human agent has empirical causal efficacy and can initiate a series as a spontaneous cause that is uncaused by any earlier event. Therefore, the idea that human choices are free and have causal efficiency with respect to natural objects of nature is built into the central hypothesis of Abellán et al [2018] and it also finds support in Allison's interpretation of Kant's theory of human free agency.

## 2. The Theorem

In a traditional Bell experiment a pair of entangled particles are spatially separated and measured by two observers, conventionally called Alice and Bob. The variables **x** and **y** are the setting variables that describe the possible measurement settings that can be performed by them respectively. The variables **a** and **b** represent the measurement outcomes of the observables associated to **x** and **y**.

Bell's theorem imposes restrictions on the probability **P(a,b|x,y)** through assumptions about the causal structure that represents the physical system in a Bell experiment. The conditions are (i) local causality, and (ii) measurement independence. "Local causality" refers to the assumption that the value of the variable **a** (**b**) is the joint effect of a hidden variable $\lambda$ and the setting variable **x** (**y**).

The experiments performed by two observers – Alice and Bob – are space-like separated events, therefore **a** and **b** are statistically independent given $\lambda$, **x** and **y**. Also, **a** (**b**) is statistically independent of **y**(**x**). Or in mathematical terms: **P(a,b|x,y,$\lambda$) = P(a|x,$\lambda$) P(b|y,$\lambda$)**.

The measurement independence condition states that the setting variables are independent of $\lambda$. It can be written as **P(x,y|$\lambda$) = P(x,y)** which is equivalent to **P($\lambda$|x,y) = P($\lambda$)**.

Using the assumptions of local causality and measurement independence one can write the conditional probability **P(a,b|x,y)** as:

$$\mathbf{P(a,b|x,y)} = \int d\lambda \ \mathbf{P(a,b|x,y,\lambda)P(\lambda|x,y)} = \int d\lambda \ \mathbf{P(a|x,\lambda)P(b|y,\lambda)P(\lambda)} \qquad (1)$$

If **P(a|x,$\lambda$)** and **P(b|y,$\lambda$)** can assume only the values 0 or 1, the hidden variable theory is deterministic. In this case, the value of the variable **a**(**b**) can be precisely known when **x**(**y**), $\lambda$ are

given. However, this is not a necessary condition for Bell's theorem. One may consider a non-deterministic causal hidden variable theory which allows that **P(a|x, λ)** and **P(b|y, λ)** assume intermediate values. The assumptions will give the same result shown in equation (1). According to Bell's theorem, therefore, deterministic and non-deterministic hidden variables theories (HVT) are equivalent. Since deterministic HVT are more restrictive, in this essay we will focus on random HVT, and consequently, the causal relations among variables are also in the domain of probabilistic natural causality.

The predictions of quantum mechanics are not compatible with the separable form of **P(a,b|x,y)** in Equation (1). This incompatibility can be experimentally tested. However, the assumptions of locality and measurement independence must be guaranteed in the experiment. It has been shown (Hall [2011]; Degorre et al [2005]; Hall [2010]; Brans [1988]) that if one of the assumptions is relaxed, it is possible to reproduce statistics of quantum states within a hidden variable model.

On the one hand, in order to ensure locality one must deal with technical issues in the experimental setup. The measurements **a** and the setting variable **x** have to be space-like separated from **b**. On the other hand, there is no agreement among physicists on the necessary conditions for ensuring the measurement independence condition (MIC). The problems involved are beyond the technical issues. Additional properties of the hidden variables are considered in experimental setups that aim to ensure MIC. As mentioned above, this situation of incomplete experimental conditions related to measurement independence is known as the "freedom-of-choice loophole" (FOCL).

The equivalence between the conditional probabilities **P(x,y|λ)=P(x,y)$** and **$P(λ|x,y)=P(λ)** given in the MIC points to two different causal relations among λ, **x** and **y**. In the first, λ cannot be the cause of the setting variables **x** and **y**. In the second, **x** and **y** cannot be the cause of λ. In this essay, we follow Abellán et al. [2018] and hold that FOCL is related only to the first relation. The second applies to the locality loophole.

Even though the three assumptions are all necessary for a Bell test, the first experimental implementations were performed without any concern for MIC. According to Aspect, Dalibard, and Roger [1982], the setting variables **x** and **y** were chosen into the backward light cone of the entangled pair creation event. This allows for the possibility of causal relations among the setting variables and λ. For Weihs et al. [1998], the setting variables were chosen in the future light cone of the emission event. In this case, **x** and **y** could have been caused by λ. In both situations, it was assumed that λ must have been created at least at the same time as the entangled pair. Thus, in these earlier experiments the FOCL was left open.

In recent Bell's inequality experiments, the authors claim that FOCL was closed. Their presuppositions can be roughly divided into two groups. In the first, Scheidla et al. [2010] and Shalm et al. [2015] presuppose that the hidden variables are generated together with the entangled particles. Thus, since the spacelike separation between the particles generation event and the choice of the measurement settings is ensured, one can as a consequence assume that λ is not a cause of **x** and **y**. In the second, Hensen et al. [2015; 2016] presuppose that the measurement setting variables **x** and **y** have no causal past.

In the first group, the assumption that λ is generated together with the entangled particles is a strong and additional assumption about hidden variables that is not present in the derivation of Bell's inequality. The problem is that this additional hypothesis imposes a kind of restriction that makes these experiments not instances of violations of Bell inequalities in a general way, as they are supposed to be, but only in a very specific way, that is, in the case in which λ is generated only in a particular moment. The problem with the second kind of experiments is that the choices of settings **x** and **y** are done by a physical device (a randomizer). Therefore, as a physical device, it is submitted to the causality principle. So this central assumption in the second kind of experiments leaves the FOCL open.

Abellán et al [2018] argue that it is not possible to close FOCL "while still requiring a physical assumption" about the choice of **x** and **y**. Following the famous argument by Bell in (Bell [2004]), they hold that FOCL can be closed only if measurement setting variables are defined by human choices. Based on this argument, they performed the Big Bell Test in five continents, and twelve laboratories, using photons, single atoms, atomic ensembles and superconducting devices. A significant part of the scientific community specializing in the field of nonlocality was involved in this test. For the first time, human choices (from 100,000 volunteers) were used to select measurement settings in a Bell test. The magnitude of this collaboration (more than a hundred authors signed a single paper) shows that the scientific community is indeed concerned about the impossibility of closing FOCL using randomizers (physical devices) or any other kind of physical object.

An important question arises: why are human choices preferable to any kind of physical process for close FOCL? Although this question is not explicitly answered in (Abellán et al [2018]), the authors explicitly say that events "requiring a physical assumption" cannot close FOCL. The Bigbell Test requires thus two strong assumptions. First, since every physical event has a causal past, one cannot ensure that λ does not belong to it. And second, since human choices are not an empirical causal process, and therefore λ cannot be a cause of them, they are the only alternative for closing FOCL.

## 3. Empirical Causality vs. Human Causality Through Freedom

In this section, we present a philosophical argument based in Kant's Critical philosophy in order to support the implicit assumption in (Abellán et al [2018]), according to which human choices are independent of $\lambda$. For this purpose, it is necessary to distinguish between the causality that regulates physical processes (which we will call merely physical causality) and the causality associated with human decisions (which we will call human causality through freedom). The first presupposes that the connection between cause and effect is constituted as a change in which "the apprehension of one thing (that which happens) follows that of the other (which precedes) in accordance with a rule" (Kant, KrV, A193/B238). Cause and effect are both merely physical events and consequently we can think in $\lambda$ as a merely physical cause of a later effect. The second presupposes that the cause of a physical effect has its origin not in a merely physical event but in the human *arbitrium*, i.e., the human will, or "power of choice" (*Willkür*) In this case, since $\lambda$ is a physical variable, it is not a part of the causal chain triggered by the human *arbitrium*. This is because the cause of a physical effect in such case is not determined by any physical variable but by a human action.

In his famous Critical program, these two kinds of causality emerge from the basic distinction adopted by Kant between nature and freedom. In Kant's theory, the term "nature" refers to the set of spatiotemporal objects of experience that are subject to rules given *a priori*. Physical causality is one of these rules. Contrary to common sense, causality in Kantian terms is not a condition of a mind independent world, but instead a condition imposed by our faculty of understanding. Therefore, physical causality is one of the rules we use to represent appearances, and not a property of a mind independent world. Experience is not an event independent of the subject but instead constituted by the subject. The revolution brought about by this new approach in the field of metaphysics was of course compared by Kant, in the Preface to the second edition or B of his *Critique of Pure Reason*, to the Copernican revolution in the natural sciences. But whereas Copernicus moved the Earth from the center of the universe to the periphery of our solar system, circling the Sun, Kant's theory moves the subject from the periphery of knowledge to the epistemic center, a place previously occupied by the object.

The second kind of causality, that which is not empirical, does not belong to nature but instead is grounded in human freedom. Contrary to the realm of nature, freedom is not conditioned by spatiotemporal objects of experience and therefore it is not restricted to the laws that strictly govern appearances. Causality through freedom, as spontaneous, is able to start a causal series without being itself caused by previous physical events. Kant establishes a cognitive boundary that limits the merely physical causality to the empirical objects which are mechanically conceived. This allows the creation of a conceptual space unrestricted by such causality. Because this conceptual space does not belong to the realm of appearances, or phenomena, Kant needs to postulate another realm, which he characterizes as "noumenal." So, causality through freedom belongs not to appearances but instead to the conceptual space of noumena. Nevertheless, causality through freedom can start a causal series in a physical space. Therefore, according to Kant, the distinction between nature as phenomenal and freedom as noumenal is necessary for guaranteeing the possibility of the second kind of causality without also denying the existence of physical causality.

When applied to the rational human agent, the distinction between nature and freedom seems to result in an inconsistency. On the one hand, the rational human agent belongs to the set of appearances because, as a human animal, it is a spatiotemporal object of experience. Therefore, it is subject to empirical causality. On the other hand, the rational human agent is taken to have the ability to initiate a causal series spontaneously, without being determined by empirical causality. To overcome this apparent inconsistency, Kant presents a detailed theory about the dual "character" of rational human agency. The rational human agent has an "empirical character," which is subject to natural causality, and also an "intelligible character," which is able to spontaneously choose the course of its actions.

Thus dual character theory applies specifically to the rational human agent's will. The empirical character of the will supports a psychological causal explanation for the agent's performance, according to in which the agent's choices and actions are, to a certain degree, predictable. The agent's desires, inclinations, and beliefs are psychological factors that allow a certain degree of predictability, hence one can take these factors to be as the empirical causes of choices and actions. This empirical notion of causality underling the relation between psychological factors and human behaviors can be thought as being not of the same kind as the physical causality but as a kind of natural causality that has a causal history. From this perspective, one can conceive in terms of a natural causality this empirical character of the will. If human choices were all defined only by this empirical character of the will, we might think that ultimately human choices would be determined by a physical variable like $\lambda$. This is because psychological causality being empirical presupposes has a causal history. In this case one cannot ensure that a physical cause, such as $\lambda$, is independent of psychological factors . If that were the case, then the experiment proposed by Abellán et al [2018] test based on empirical character of human choice would not guarantee the closing of the loophole (FOCL).

In Kantian terms, the causality associated with the empirical character is not sufficient to determine the will of the rational agent. According to Kant, the choice is a deliberative process of reasoning in which the association of ideas and concepts that guide the human action requires a degree of independence from empirical causes. This presupposes a spontaneity of reason for starting a causal series which, in turn, was not determined by any previous natural cause. In this sense, it is necessary to assume a freedom of the rational agent in determining his choices and actions triggered by these choices. Kant characterizes this free character of the human will, which is independent of empirical causality, as "intelligible." By means of the intelligible character of the will, one can explain the choices and acts of a rational agent as effects of a different kind of cause that is not itself subject to the rules of empirical causality, whether merely physical or empirically psychological. Kant call this "causality through freedom."

Henry Allison uses what he calls the "Incorporation Thesis" to support the presence of the dual character of the will in Kant's philosophy. According to Allison's interpretation of Kant's theory of agency, the choices and actions of a rational agent are yielded by the incorporation of motives, like desires, inclinations, moral principles, etc., as reasons. The motives are presented to the rational human agent, who deliberately, self-consciously, and spontaneously decides to incorporate them as reasons or choice and action. The incorporation thesis allows us to identify the human power of choice as "*arbitrium liberum*" and not as "*brutum*" (Kant, KrV, A535/B562). In the animal power of choice, based on *arbitrium brutum*, sensible stimuli are causes of actions. If the human power of choice were an *arbitrium brutum*, then the choices and actions of a rational human agent would be caused by empirical psychological factors. But, for Kant, this is not the case. In Allison's words, "such a subject is, therefore, more properly characterized as a patient rather than an agent". (Allison [1996], p.130). To think of human rational agency as an *arbitrium liberum* means that inclinations and desires are not direct causes of actions. The rational human agent deliberately and spontaneously decides to incorporate motives, even though the agent is aware of desires and inclinations. Being aware of desires and inclinations does not means that they have a causal function in the deliberative and decision-making process. The agent can freely incorporate or not incorporate a motive as a maxim or a rule and, and choose or act according to them.  Allison understands the act of incorporation as a genuine causal factor for any rational human action (Allison [1990]). Thus, the incorporation is an efficient cause of the choice or action for a rational human agent. Therefore, it must take place in a certain period of time which starts a causal series. But the incorporation itself is not the result of previous causes. It initiates by itself (as a spontaneous or uncaused cause) a new causal series. This type of causality, causality through freedom, can be thought only as an intelligible character of the human power of choice.

We must emphasize that the Incorporation Thesis is based on the absolute spontaneity of the *arbitrium*, without any empirical causal influence. Therefore, it requires an absolute kind of freedom, which does not presuppose any previous merely physical or empirically psychological cause, which Kant calls "transcendental freedom." The incorporation of a motive as a rule for rational human choice and action starts a new causal series. This condition of absolute freedom of the agent is more restrictive then the condition required to the solution of FOCL in Bell experiments with human choices. As was shown by Hall [2010], the complete independence between the hidden variables **λ** and settings variables **x** and **y** is not required. The level of tolerance for the measurement independence condition is low (indeed, it was even calculated by Hall [2010]) but it is not null. Therefore, when one accepts that the settings variables are defined by human choices, these choices would not necessarily have to be absolutely free. One could consider a degree (very low) of dependence between the hidden variables and human choices and still close the FOCL. In this sense, the specifically Kantian position on the freedom of rational human agents is more restrictive then the condition required to close FOCL with human choices.

As regards merely physical causality, the Kantian position is also more restrictive than Bell's position. Kant holds that merely physical causality is deterministic, while in Bell's theorem the causality can be taken as probabilistic. Kant's theory of rational human agency requires the compatibility of absolute noumenal freedom and merely physical phenomenal causal determinism. These Kantian conditions are stronger than those associated with the application of human choice to close FOCL. In the latter, what is required is compatibility of a quasi-absolute freedom (as demonstrated in the works of Hall [2011]; Degorre et al [2005]; Hall [2010]) and probabilistic or indeterministic causality. Therefore, Kant's theory of agency is better suited to support the free will assumption in (Abellán et al [2018]), because it is able to resolve more serious inconsistencies.

The Kantian solution that allows the compatibility between absolute noumenal freedom and merely physical phenomenal causal determinism cannot be easily classified as either "compatibilist" or "incompatibilist" in the traditional sense. This is a matter of debate among interpreters (Hanna [2006]; Wood [1984]; Allison [1990]). The peculiarity of Kantian position is noted by Robert Hanna as follows: "Kant's libertarian theory of freedom of the will is philosophically significant precisely because it is neither hard determinist, nor soft determinist, nor causal indeterminist, nor hard indeterminist, nor compatibilist, nor incompatibilist." (Hanna, [2006], p.419). This because, on the one hand, the absolute noumenal freedom required by Kant's solution is not compatible with deterministic merely physical phenomenal causality, and, on the other hand, the introduction of a conceptual space—i.e., noumenal space--free from empirical causality allows Kant to claim that causality through freedom is not in direct causal competition with empirical causality. In this sense, nature and freedom are not in direct opposition but can be thought as yielding causal series which are compatible but not in any way reducible or otherwise assimilable to each other. Therefore, this view is sometimes called Kant's "incompatibilistic compatibilism."

In conclusion, we argue that the alternative for closing the FOCL through the hypothesis of free will can find philosophical support in Kant's theory of rational human agency. By promoting an "incompatibilistically compatibilist" approach to the relation between freedom and nature, Kant's theory provides philosophical support for the central hypothesis of the experiment carried out in (Abellán et al. [2018]). The requirement that the capacity of the rational human agent for choosing the directions of the measurement setup free from any causal influence, is fulfilled by Kant's theory of rational human agency. According to Allison's interpretation, the incorporation of a motive as a rule for choice and an action has efficient causal power even though it is not merely physically or empirically psychologically caused by previous events. In other words, it has no phenomenal causal history. It is a spontaneous cause that starts a new series of phenomenal events. This kind of causality through freedom is exactly what validates the use of human choices, as free choices, to decide the directions of the measures in the experiment. Therefore, his Kantian philosophical point of view enables us to conclude that rational human agency is the right candidate for closing the FOCL.

**Acknowledges**

We would like to thank Robert Hanna for his critical comments and helpful suggestions.

*Department of Philosophy*
*Universidade Federal de Minas Gerais*
*Belo Horizonte, Brazil*

*Instituto de Ciências Exatas e Tecnológicas*
*Universidade Federal de Viçosa*
*Florestal, Brazil*

*romeu.rossi@gmail.com*
*pkauark@gmail.com*